\documentclass[twocolumn,showpacs,aps,prd,nobibnotes,nofootinbib,floatfix]{revtex4-1}

\usepackage{amsmath}
\usepackage{graphicx,subfigure}
\usepackage[usenames]{color}
\usepackage[normalem]{ulem}
\usepackage{textcomp}
\usepackage{gensymb}
\usepackage{xspace}
\usepackage{verbatim}

\usepackage{soul} 

\usepackage[colorlinks=true]{hyperref}
\hypersetup{
  citecolor  = blue
}

\begin{document}

\title{Weak cosmic censorship conjecture for the novel $4D$ charged Einstein-Gauss-Bonnet black hole with test scalar field and particle}

\author{Si-Jiang Yang}
\author{Jun-Jie Wan}
\author{Jing Chen}
\author{Jie Yang}%
\email{yangjiev@lzu.edu.cn}

\author{Yong-Qiang Wang}%
\email{yqwang@lzu.edu.cn, corresponding author}

\affiliation{Institute of Theoretical Physics $\&$ Research Center of Gravitation, Lanzhou University, Lanzhou 730000, China}

\date{\today}

\begin{abstract}
  Recent researches of the novel $4D$ Einstein-Gauss-Bonnet (EGB) gravity have attracted great attention. In this paper, we investigate the validity of the weak cosmic censorship conjecture for a novel $4D$ charged EGB black hole with test charged scalar field and test charged particle respectively. For the test charged field scattering process, we find that both extremal and near-extremal black holes cannot be overcharged. For the test charged particle injection, to first order, an extremal black hole cannot be overcharged while a near-extremal $4D$ charged EGB black hole can be destroyed. To second order, however, both extremal and near-extremal $4D$ charged EGB black holes can be overcharged for positive Gauss-Bonnet coupling constant; for negative Gauss-Bonnet coupling constant, an extremal black hole cannot be overcharged and the validity of the weak cosmic censorship conjecture for a near-extremal black hole depends on the Gauss-Bonnet coupling constant.
\end{abstract}

\maketitle

\section{Introduction}
\label{sec:intro}

It is well known that the EGB gravity is one of the most promising candidates for modified gravity. However, in four-dimensions, the Gauss-Bonnet term is a topological invariant and has no contribution to the field equation, which results in trivial black hole solutions in four dimensions. Recently, by rescaling the Gauss-Bonnet coupling parameter $\alpha\rightarrow\alpha/(D-4)$ and then taking the limit $D\rightarrow 4$, Glavan and Lin obtained a general covariant EGB modified theory of gravity in four-dimensions and presented a novel vacuum black hole solution \cite{GlLi20,CMSZ13}. The novel $4D$ EGB gravity bypasses the Lovelock's theorem and avoids Ostrogradsky instability.
Thus, the Gauss-Bonnet term has nontrivial contribution to the gravitational field equations. Following the spirit of Glavan and Lin, Fernandes generalized their black hole solution to include electric charge and got a $4D$ charged spherical solution \cite{Fern20}. Later, applying Newman-Janis algorithm, Wei and Liu et al. obtained a $4D$ rotating EGB black hole solution \cite{WeLi20b,KuGh20}.

There are a lot of following papers inspired by the research of Glavan and Lin, such as the shadow and inner most stable circular orbits of black hole in the EGB gravity \cite{WeLi20b,KoZi20, GuLi20,RoCh20}, thin accretion disk around four-dimensional EGB black holes \cite{LiZW20}, spinning test particle in four-dimensional EGB black hole \cite{ZhWL20}, gravitational lensing and bending of light \cite{JiGL20,HeHS20,IsKG20}, black hole as particle accelerator \cite{KRHAM20}, black hole thermodynamics and phase transition \cite{WeLi20a,HKRAA20,SiSi20,Mans20,EsJa20}, quasinormal modes and strong cosmic censorship \cite{Mish20,Chur20,Aragon:2020qdc}, scalar field in $4D$ EGB black holes \cite{ZZLG20}, cosmology in the EGB gravity \cite{SLli:2020}, compact objects and their  properties \cite{DoYa20}, the eikonal gravitational instability
of asymptotically flat and (A)dS black holes \cite{Konoplya:2020juj}, greybody factor and power spectra of the Hawking radiation \cite{ZhLG20,KoZi20b}, other black hole solutions \cite{LuPa20,GhKu20,KuKu20,KoZh20, CCRV20,KuGh20a,SiGM20,GhMa20a}, and related topics about higher-derivative gravity in two dimensions \cite{NoOd20,Ai20}.

The weak cosmic censorship conjecture states that spacetime singularities arising in gravitational collapse should always been hidden behind black hole event horizons \cite{Penr69}.
Although lacking a general proof, the weak cosmic censorship conjecture has become one of the cornerstones for black hole physics. To test the validity of Penrose' weak cosmic censorship conjecture, Wald proposed a gedanken experiment to check the validity of the weak cosmic censorship conjecture by throwing a particle with large charge or angular momentum into an extremal Kerr-Newman black hole \cite{Wald74}. The result demonstrates that particles causing the destruction of the event horizon would not be captured by the black hole. Later,  Hubeny's pioneer work shows that the horizon of a near-extremal charged black hole can be destroyed by dropping a charged test particle \cite{Hube99}, the result is the same for a near-extremal Kerr black hole \cite{JaSo09}. Recently,  Sorce and Wald proposed a new version
of the thought experiment to destroy a nearly extremal Kerr-Newman
black hole at the second-order approximation of the perturbation that comes from the matter fields \cite{SoWa17}, the result shows that the event horizon of the Kerr-Newman black hole cannot be destroyed. Using the method of Sorce and Wald, the systematic works of Jiang et al. suggest that a series of black holes cannot be destroyed \cite{JiZh20,GMZZ18,NiCL19}. Another way of destroying the event horizon of a black hole to test the validity of the weak cosmic censorship conjecture is the scattering of a test classical field first proposed by Semiz \cite{Semi11}, and further developed by others \cite{DuSe13,SeDu15,Duzt15}. Recently, Gwak divided the scattering process into a series of infinitesimal time interval and considered an infinitesimal process only, the result shows that Kerr-(anti) de Sitter black holes cannot be overspun by a test scalar field \cite{Gwak18}, and many works have been done following this line \cite{Gwak19a,Gwak19b,Chen18,ChZY19,ZeHC19}. The dividing of infinitesimal time interval process may provide clues that time interval for particles crossing the event horizon may be important for considering the weak cosmic censorship conjecture \cite{YCWWL20,LiWL19,Gwak17}. For other works to destroy the horizon of a black hole, see Refs. \cite{NaQV16,BaCK10,RoCa11,LCNR10,ShDA19,Hod02,ZVPH13,CoBa15, Goncalves:2020ccm,BaCK11,CBSM15,MaSi07,Hod08,RiSa08, MRSSV09,RiSa11,LiGa20,Han:2019lfs,WaWY20}. For a brief review on weak cosmic censorship conjecture with some thoughts see Ref. \cite{Ong20}.

The weak cosmic censorship conjecture has become one of the cornerstones of black hole physics, and it has deep connection with the laws of black hole thermodynamics. The conjecture might be proved to be true, but the gedanken experiment
%to investigate the validity of the weak cosmic censorship conjecture
can help us to understand it better and bring new insight on the relation between the weak cosmic censorship conjecture and the laws of black hole thermodynamics. Besides, it is also important to explore what is it that makes the conjecture to be true. On the other hand, if the final result proves that the weak cosmic censorship conjecture can indeed be violated, this might provide us the possibility to access regions of high curvature and provide us observable information to build a consistent theory of quantum gravity \cite{LiBa13}.

Recent researches of weak cosmic censorship conjecture in extended phase space have attracted a lot of attention, but also aroused controversy. The disagreement primarily focuses on the following point \cite{Gwak:2017kkt}: when a particle with energy $E$ is dropped into a black hole in AdS space, the internal energy of the black hole $U\equiv M-PV$ increases by the particle energy, i.e., after absorption of the particle, the internal energy of the black hole is $U'=U+E$ instead of enthalpy $M'=M+E$. The crucial flaw of this viewpoint is that it would lead to the violation of the second law of black hole thermodynamics.
The viewpoint was opposed by Page et al.. They argued that when a particle is dropped into a black hole, it is the enthalpy that increases by the particle energy instead of the internal energy \cite{HuOP19}. The argument of Page et al. preserves the second law of black hole thermodynamics naturally.

In this paper, we investigate the weak cosmic censorship conjecture of a novel $4D$ charged EGB black hole by the scattering of a massive charged scalar field and a charged test particle in the normal phase space, respectively. For scattering of a charged scalar field, our result suggests that both the extremal and near-extremal charged EGB black holes cannot be overcharged. For test particle injection, the study suggests that to first order an extremal $4D$ charged EGB black hole cannot be overcharged and a near-extremal black hole can be destroyed. However, to second order the validity of the weak cosmic censorship conjecture depends on the value of the Gauss-Bonnet coupling constant. %By checking the validity of the weak cosmic censorship conjecture, we can get constraints on the Gauss-Bonnet coupling constant from the validity of the weak cosmic censorship conjecture.

The structure of the paper is as follows. In Sec.\,\ref{2}, we briefly review the $4D$ charged EGB black hole and its thermodynamics. In Sec.\,\ref{3}, we explore the scattering of a massive complex scalar field in the $4D$ charged EGB black hole background and obtain the energy and charge fluxes of the complex scalar field. In Sec.\,\ref{4}, we check the validity of the weak cosmic censorship conjecture for extremal and near-extremal black holes by scattering of the charged scalar field. In Sec.\,\ref{5}, we check the possibility of overcharging the black hole by injection of a test charged particle. We compare our research with other works in Sec.\,\ref{6}. The last section is devoted to discussions and conclusions.

\section{The novel $4D$ charged EGB black hole and its thermodynamics}\label{2}

The action of the EGB gravity with electromagnetic field in a $D$-dimensional spacetime is
\begin{eqnarray}
\begin{split}
 &&S=\frac{1}{16\pi}\int d^Dx\left(R+\frac{\alpha}{D-4}L_{GB}-F_{\mu\nu}F^{\mu\nu}\right),
 \end{split}
 \end{eqnarray}
with the Gauss-Bonnet term
 \begin{eqnarray}
 &&L_{GB}=R_{\mu\nu\rho\sigma}R^{\mu\nu\rho\sigma}-4R_{\mu\nu}R^{\mu\nu}+R^2,
\end{eqnarray}
where $\alpha$ is the Gauss-Bonnet coupling constant, and $F_{\mu\nu}$ is the electromagnetic field strength tensor $F_{\mu\nu}=\partial_\mu A_\nu-\partial_\nu A_\mu $, with $A_\mu$ being the electromagnetic vector potential of the black hole.

In the effective action approach of string theory, the Gauss-Bonnet term is the leading order quantum correction to gravity \cite{GrSl87}. The Gauss-Bonnet coupling constant is related to string scale and can be identified with the inverse of string tension. Besides, the Gauss-Bonnet coupling constant indicates the leading quantum gravity correction from string theory \cite{ChDu02}.

The static spherical charged black hole solution was obtained by solving the field equations in D-dimensional spacetime and taking the limit $D\rightarrow4$ in Ref. \cite{Fern20}:
\begin{eqnarray}
\begin{split}
 ds^2&=&-f(r)dt^2+\frac{1}{f(r)}dr^2+r^2(d\theta^2+\sin^2\theta d\phi^2),\\
 f(r)&=&1+\frac{r^2}{2\alpha}\left[1-\sqrt{1+ 4\alpha\left(\frac{2M}{r^3}-\frac{Q^2}{r^4}\right)}\right],
 \end{split}
\end{eqnarray}
with the electromagnetic 4-vector potential $A=-Q/r \, \, \, \text{d}t$. The parameters $M$ and $Q$ are the mass and charge of the black hole, respectively. Taking the limit $\alpha\rightarrow 0$, the metric reduce to the Reissner-Nordstr\"om black hole solution. The black hole solution has the same form with the one obtained in a comformal anomaly gravity \cite{CaCO10,Cai14}. This spacetime is singular at $r=0$ due to the divergence of curvature scalar.

The metric function can be written as
%\begin{eqnarray}
%% \nonumber to remove numbering (before each equation)
%  \nonumber f(r) &=& \frac{2(r^2-2Mr+Q^2+\alpha)}{r^2+2\alpha+\sqrt{r^4+4\alpha(2Mr-Q^2)}} \\
%   &=& \frac{2\Delta}{r^2+2\alpha+\sqrt{r^4+4\alpha(2Mr-Q^2)}},
%\end{eqnarray}
\begin{equation}
\begin{split}
  f(r)=&\frac{2(r^2-2Mr+Q^2+\alpha)}{r^2+2\alpha+ \sqrt{r^4+4\alpha(2Mr-Q^2)}} \\ =&\frac{2\Delta}{r^2+2\alpha+\sqrt{r^4+4\alpha(2Mr-Q^2)}},
\end{split}
\end{equation}
where we have defined
\begin{equation}
  \Delta=r^2-2Mr+Q^2+\alpha.
\end{equation}
The event horizon is determined by the equation $f(r)=0$, which is equivalent to the following equation
\begin{equation}
  \Delta=r^2-2Mr+Q^2+\alpha=0. \label{Hequation}
\end{equation}
For a non-extremal charged EGB black hole, the above equation gives the inner and outer horizons
\begin{equation}\label{horiz}
    r_{\pm}=M\pm \sqrt{M^2-Q^2-\alpha}.
\end{equation}
The event horizon is the outer horizon $r_+$. The two horizons coincide for an extremal black hole, and the degenerate horizon locates at $r_{\text{ex}}=M$. For $M^2<Q^2+\alpha$, the horizon disappears and there is no black hole. In this case, the metric describes a charged naked singularity. For convenience, we denote the event horizon $r_+$ as $r_{\text{h}}$ in the following.

The temperature of the black hole can be calculated as
\begin{equation}
  T=\frac{r_{\text{h}}^2-Q^2-\alpha}{8\pi\alpha r_{\text{h}} +4\pi r_{\text{h}}^3}.\label{Tempe}
\end{equation}
The area of the event horizon of the black hole is
\begin{equation}
  A=4\pi r_{\text{h}}^2, \label{area}
\end{equation}
and the electric potential of the event horizon is
\begin{equation}\label{epotential}
  \phi_{\text{h}}=\frac{Q}{r_{\text{h}}}.
\end{equation}
The first law of thermodynamics for the black hole is \cite{WeLi20a}
\begin{equation}
dM=TdS+\phi_{\text{h}}dQ+\mathcal{A}d\alpha, \label{1stlaw}
\end{equation}
with the entropy and the conjugate quantity $\mathcal{A}$ to the coupling parameter $\alpha$
\begin{equation}
\begin{split}
  S=\pi r^2_{\text{h}}+4\pi\alpha \ln \left(\frac{r_{\text{h}}}{\sqrt{|\alpha|}}\right),
  \end{split}
\end{equation}
\begin{equation}
\begin{split}
  \mathcal{A}=&\left(\frac{\partial M}{\partial \alpha}\right)_{S, Q}\\=&\frac{\alpha +2 \ln
   \left(\frac{r_{\text{h}}}{\sqrt{|\alpha|
   }}\right) \left(\alpha -r_{\text{h}}^2+Q^2\right)+2 r_{\text{h}}^2-Q^2}{2 \left(2 \alpha
   r_{\text{h}}+r_{\text{h}}^3\right)}.
   \end{split}
\end{equation}

It is worth noting that the entropy of the $4D$ charged EGB black hole is different from the usual Bekenstein-Hawking entropy-area law due to the existence of the Gauss-Bonnet coupling constant $\alpha$, and this entropy is consistent with that obtained from the Iyer-Wald formula \cite{WeLi20a,LuPa20}.

\section{Charged massive scalar field in charged EGB space-time}\label{3}

\subsection{The scattering for charged massive scalar field}

We consider the scattering of charged massive scalar field in the $4D$ charged EGB spacetime background. The charged massive scalar field $\Psi$ with mass $\mu_{\text{s}}$ and charge $q$ minimally coupled to the gravity is governed by the equation of motion
\begin{equation}
  (\nabla_\mu-iqA_\mu)(\nabla^\mu-iqA^\mu)\Psi-\mu_{\text{s}}^2\Psi=0,
\end{equation}
which can be written as
\begin{equation}\label{field}
\begin{split}
   \frac{1}{\sqrt{-g}}(\partial_\mu-iqA_\mu)\left[\sqrt{-g}g^{\mu\nu}(\partial_\nu -iqA_\nu) \Psi\right]-\mu_{\text{s}}^2\Psi=0.
   \end{split}
\end{equation}

Since the spacetime is static and spherically symmetric, the complex scalar field can be decomposed into the following form \cite{Teuk72,Teuk73}
\begin{equation}
   \Psi(t,r,\theta,\phi)=e^{-i\omega t} R_{lm}(r)Y_{lm}(\theta,\phi),\label{wavefun}
\end{equation}
where $ Y_{lm}(\theta,\phi) $ are spherical harmonic functions and $R_{lm}(r)$ are the radial functions.
Inserting the above equation into the equation of motion Eq. (\ref{field}), we get the equation for the radial part
\begin{equation}
\begin{split}
   \frac{1}{r^2}\frac{d}{dr}\left[r^2f(r)\frac{dR_{lm}}{dr}\right]&+ \left[\frac{(\omega-\frac{qQ}{r})}{f(r)} \right. \\ &\left.-\frac{l(l+1)}{r^2}
   -\mu_{\text{s}}\right]R_{lm}=0,
   \end{split} \label{radial}
\end{equation}
and angular part
\begin{equation}
\begin{split}
   \left[\frac{1}{\sin\theta}\frac{\partial}{\partial\theta}\left( \sin\theta\frac{\partial}{\partial\theta}\right)+ \frac{1}{\sin^2 \theta} \frac{\partial^2}{\partial\phi^2}\right]Y_{lm}=-l(l+1)Y_{lm},
   \end{split} \label{angu}
\end{equation}
where $l(l+1)$ is the separation constant and $l$ takes positive integra values.
The solutions to the angular part of the equation are the spherical harmonic functions. Since the angular solution is well known and it can be normalized to unity, we are more interested in the radial part.

To solve the radial equation, we introduce the tortoise coordinate as usual
\begin{equation}\label{tortoise}
   \frac{dr}{dr_*}=f(r).
\end{equation}
Then, the radial equation has the following form
\begin{equation}
\begin{split}
     \frac{d^2R_{lm}}{dr_*^2}&+\frac{2f(r)}{r}\frac{dR_{lm}}{dr_*}+ \left[(\omega-\frac{qQ}{r})^2\right. \\ &\left.-f(r)\left(\frac{l(l+1)} {r^2}-\mu_{\text{s}}^2\right) \right]R_{lm} =0.
     \end{split} \label{rad}
\end{equation}
when $r$ varies from the horizon $r_{\text{h}}$ to infinity, the tortoise coordinate ranges from $-\infty $ to $+\infty$ , and thus covers the whole space outside the event horizon.

It is convenient to investigate the radial equation near the horizon since we are more concerned with waves incident into the black hole. Near the horizon, Eq. (\ref{rad}) can be approximated as
\begin{equation}\label{radia}
   \frac{d^2R_{lm}}{dr_*^2}+ \left(\omega-\frac{qQ}{r_\text{h}}\right)^2R_{lm}=0.
\end{equation}
Using Eq.~(\ref{epotential}), the above equation can be written as the following form
\begin{equation}\label{radiala}
   \frac{d^2R_{lm}}{dr_*^2}+ \left(\omega-q\phi_\text{h}\right)^2R_{lm}=0.
\end{equation}
The solution of the above radial equation is
\begin{equation}\label{radaialsol}
   R_{lm}(r)\sim \exp[\pm i(\omega-m\phi_\text{h})r_*].
\end{equation}
The positive sign corresponds to outgoing wave modes; while, the negative sign corresponds to the ingoing wave modes. We choose the negative sign since the ingoing wave mode is the physically acceptable solution. Thus, the charged complex scalar field near the event horizon has the form
\begin{equation}\label{sol}
   \Psi=\exp[-i(\omega-q\phi_\text{h})r_*]Y_{lm}(\theta,\phi)e^{-i\omega t}.
\end{equation}

After obtaining the wave function, we can calculate the parameter changes of the black hole through the energy momentum and charge flux of the complex scalar field.

\subsection{Thermodynamics during scattering of the charged scalar field}

Since the $4D$ charged EGB black hole is non-rotating, we shoot a single wave mode $(l, m=0)$  into the black hole. The parameter changes of the black hole can be estimated from the fluxes of the charged scalar field during the scattering. The energy-momentum tensor of the charged scalar field is given by
\begin{equation}\label{energy-momentun}
\begin{split}
   T^\mu_\nu=&\frac{1}{2}\mathcal{D}^\mu\Psi\partial_\nu\Psi^*+ \frac{1}{2}\mathcal{D}^{*\mu}\Psi^*\partial_\nu\Psi \\
   &-\delta^\mu_\nu(\frac{1}{2} \mathcal{D}_\alpha\Psi\mathcal{D}^{*\alpha}\Psi^*- \frac{1}{2}\mu_{\text{s}}\Psi\Psi^*),
   \end{split}
\end{equation}
with
\begin{equation}
  \mathcal{D}=\partial_\mu-iqA_\mu.
\end{equation}
From Eq.~(\ref{energy-momentun}), it is easy to get the energy flux through the event horizon %\cite{Padm10}
\begin{equation}\label{Eflux}
   \frac{dE}{dt}=\int_{\text{H}} T^r_t\sqrt{-g} \, d\theta d\phi=\omega(\omega-q\phi_\text{h})r_\text{h}^2.
\end{equation}
The electric current of the charged scalar field is
\begin{equation}\label{ecurrent}
  j^\mu=-\frac{1}{2}iq(\Psi^*\mathcal{D}^{\mu}\Psi-\Psi \mathcal{D}^{*\mu}\Psi^*).
\end{equation}
The charge flux through the event horizon is
\begin{equation}
   \frac{dQ}{dt}=-\int_{\text{H}} j^r\sqrt{-g}d\theta d\phi=q(\omega-q\phi_\text{h})r_\text{h}^2.
\end{equation}
Where we have used the normalization condition for the spherical harmonic functions $Y_{lm}(\theta,\phi)$ in the integration. The ratio of the charge flux to the energy flux is $q/\omega$ as indicated in Ref. \cite{Beke73}.

From the fluxes of the energy and charge, it is clear that the energy and charge flow into the black hole for wave modes with $\omega > q \phi_\text{h}$; while, the energy and charge fluxes are negative for wave modes with $\omega < q\phi_\text{h}$, which indicates that the scalar field extract energy and charge from the black hole. This is called black hole superradiance \cite{BrCP15}.

Consider an infinitesimal time interval $dt$, the changes in the mass and charge of the black hole are
\begin{eqnarray}
 % \nonumber to remove numbering (before each equation)
   dM&=&dE = \omega(\omega-q\phi_\text{h})r_\text{h}^2 \, dt , \label{Echange} \\
   dQ &=&  q(\omega-q\phi_\text{h})r_\text{h}^2 \, dt. \label{qchange}
\end{eqnarray}

If we consider a black hole far from extremal, the final state is still a black hole after the absorption of the infinitesimal energy and charge of the complex scalar field. The change in the black hole configuration can be represented in terms of the frequency  $\omega$ and charge $q$ of the complex scalar field. The change in the location of the horizon $dr_{\text{h}}$ can be obtained from the condition
\begin{equation}
\begin{split}
     \Delta(M+dM, Q+dQ, r_{\text{h}}&+dr_{\text{h}})=\frac{\partial\Delta}{\partial M}dM \\
     &+\frac{\partial\Delta}{\partial Q}dQ+\frac{\partial\Delta}{\partial r_{\text{h}}}dr_{\text{h}}=0.
     \end{split}
\end{equation}
Then, we obtain the change of the horizon for the scattering process
\begin{equation}\label{rhchange}
     dr_{\text{h}}=\frac{r_{\text{h}}dM-QdQ}{r_{\text{h}}-M}.
\end{equation}
The change of the black hole area is
\begin{equation}\label{areaincrease}
      dA=\frac{8\pi r_{\text{h}}^2}{\sqrt{M^2-Q^2-\alpha}}\, (\omega-q\phi_{\text{h}})^2\, \, dt,
\end{equation}
which is always positive. This indicates that the area of the event horizon never decreases during the scattering of the complex scalar field, and it is consistent with Hawking's area increasing theorem, which states that the area of a black hole event horizon never decreases during classical process  \cite{Hawk72,CCFK18}.

\section{Overcharging the black hole with the charged massive scalar field}\label{4}

In this section, we investigate whether extremal and near-extremal $4D$ charged EGB black holes can be destroyed by the charged scalar field during the scattering. We shoot a monotonic classical test complex scalar field with frequency $\omega$ and azimuthal harmonic index $m=0$ into the extremal or near-extremal charged EGB black holes. To examine whether we can overcharge the black hole, we only need to check the existence of event horizon after the scattering.

The metric function $\Delta$ determines the black hole event horizon
\begin{equation}
  \Delta=r^2-2Mr+Q^2+\alpha.
\end{equation}
The metric function $\Delta$ takes the minimal value at the point $r_{\text{min}}=M$ with
\begin{eqnarray}
\Delta_{\text{min}}=Q^2+\alpha-M^2. \label{Deltamin}
\end{eqnarray}
If the minimum of the metric function is negative or zero, the metric describes a black hole; while, if the minimal value of $\Delta$ is positive, there is no black hole.

It is convenient for us to consider a small time interval $dt$. For the whole scattering process, we can divide it into a series of small time intervals and consider each intervals separately by only changing the parameters of the black hole.

In the the small time interval $dt$, the black hole absorbs the complex scalar field with energy $dE$ and charge $dQ$. The change of the black hole parameters are
\begin{eqnarray}
M &\rightarrow&  M'=M+dM,~~~~\nonumber\\
Q &\rightarrow& Q'=Q+dQ, ~~~~\\
\alpha &\rightarrow& \alpha'=\alpha.\nonumber
\end{eqnarray}

When the fluxes of the charged scalar field enter into the black hole, the minimum of the metric function $\Delta_{\text{min}}$ changes to $\Delta'_{\text{min}}$,
\begin{eqnarray}
% \nonumber to remove numbering (before each equation)
\nonumber  \Delta'_{\text{min}} &=& \Delta'_{\text{min}}(M+dM, Q+dQ, \alpha) \\
\nonumber   &=& \Delta_{\text{min}} +\left(\frac{\partial\Delta_{\text{min}}}{\partial M}\right)_{Q,\alpha}dM+ \left(\frac{\partial\Delta_{\text{min}}}{\partial Q}\right)_{M,\alpha}dQ\\
   &=&  -(M^2-Q^2-\alpha)+2QdQ-2MdM.
    \label{Delta'min1}
\end{eqnarray}

To check the validity of the weak cosmic censorship conjecture, we assume the black hole starts out extremal or very close to extremal. Now, the question is whether the metric function $\Delta=0 $ has a positive solution after the black hole absorbs the test charged scalar field during the scattering, or equivalently, whether the minimum $\Delta_{\text{min}}$ of the metric function is positive after the absorption of the charged scalar field.

For a near-extremal black hole, the event horizon radius $r_{\text{h}}$ is extremely close to the point of minimal value $r_{\text{min}}=M$, we define an infinitesimal distance $\epsilon$ between the event horizon $r_{\text{h}}$ and the minimal point $r_{\text{min}}$:
\begin{equation}
 r_\text{h}=r_{\text{min}}+\epsilon. \label{distance}
\end{equation}
It is clear that $\epsilon >0$ describes a near-extremal black hole and $\epsilon=0$ correspond to the extremal black hole. Before the absorption of the scalar field, we can write the minimum of the metric function $\Delta$ as
\begin{equation}
   \Delta_{\text{min}}=Q^2+\alpha-M^2=-\epsilon^2. \label{iminimal}
\end{equation}

During the infinitesimal time interval $dt$ of  the scattering, the black hole absorbs the charged scalar field and the minimum of the metric function $\Delta_{\text{min}}$ becomes $\Delta'_{\text{min}}$:
\begin{equation}
     \Delta'_{\text{min}}= -(M^2-Q^2-\alpha)-2MdM +2QdQ. \label{HChecking}
\end{equation}
Plugging Eq. (\ref{Echange}) and Eq. (\ref{qchange}) into the above equation and to first order in $dt$, we have
\begin{equation}
   \Delta'_{\text{min}}= -\epsilon^2-2Mq^2(\frac{\omega}{q}-\phi_{\text{h}}) (\frac{\omega}{q}-\frac{Q}{M})r_{\text{h}}^2 \, dt,  \label{Delta'min}
\end{equation}
where $\phi_\text{h}$ is the electric potential of the black hole as defined in Eq.~(\ref{epotential}).

To check the validity of the weak cosmic censorship conjecture for the $4D$ charged EGB black hole, we consider extremal black hole first. For extremal $4D$ charged EGB black hole, we have
\begin{equation}
  M^2-Q^2-\alpha=0,
\end{equation}
and the electric potential
\begin{equation}
  \phi_{\text{h}}=\frac{Q}{M}.
\end{equation}
Then, after the absorption of the charged scalar field, the minimal value of the metric function is
\begin{equation}
 \Delta'_{\text{min}}=-2Mq^2(\frac{\omega}{q} -\phi_{\text{h}})^2r_{\text{h}}^2dt, \label{extrmin}
\end{equation}
which is always negative. It is clear that the extremal black hole becomes non-extremal and has two event horizons after absorption of scalar field with $\omega\neq q\phi_{\text{h}}$. While for charged scalar field with $\omega=q\phi$, the extremal black hole will still be extremal since the scalar field neither be absorbed nor extract energy from the black hole. This suggests that the extremal black hole cannot be overcharged, and the weak cosmic censorship conjecture is valid for extremal $4D$ EGB black hole.

For a near-extremal $4D$ EGB black hole, we only need to check whether the charged scalar field with mode
\begin{equation*}
  \frac{\omega_0}{q}=\frac{1}{2}(\frac{Q}{M}+ \phi_{\text{h}})
\end{equation*}
can destroy the event horizon. Since scalar field belongs to this mode, the metric function $\Delta'_{\text{min}}$ is the largest. Thus, the near-extremal black hole can be destroyed unless this mode of the scalar field can overcharge the black hole.

We shoot a wave with the mode $(\omega_0, m=0) $ into the near-extremal $4D$ charged EGB black hole. Then, the minimal value of the metric function is
\begin{equation}
  \Delta'_{\text{min}}=-\epsilon^2+\frac{q^2Q^2}{2M}\epsilon^2dt. \label{Deta'm}
\end{equation}
Since the time interval is infinitesimal and $dt\sim \epsilon$. Then, we have
\begin{equation}
  \Delta'_{\text{min}}=-\epsilon^2+ \frac{q^2Q^2}{2M}\epsilon^3<0. \label{Deltamin'}
\end{equation}
Which shows that the near-extremal $4D$ charged EGB black hole always has two event horizons after the scattering. This indicates that near-extremal $4D$ EGB black hole cannot be overcharged and the weak cosmic censorship conjecture is valid.

Thus, both extremal and near-extremal $4D$ charged EGB black holes cannot be overcharged by the charged scalar field during the scattering and the weak cosmic censorship conjecture is valid. It is worth noting that for charged scalar field injection the Gauss-Bonnet coupling constant has no effect on the validity of the weak cosmic censorship conjecture. The weak cosmic censorship conjecture is still valid for both positive and negative coupling constant, and for vanishing Gauss-Bonnet coupling constant, the result recovers to that of Reissner-Nordstr\"om black hole.

\section{Overcharging the black hole with test particle}\label{5}

Now we attempt to destroy the event horizon of the $4D$ charged EGB black hole with a test charged particle. The gedanken experiment to destroy the event horizon of an extremal black hole with large charge or large angular momentum was first designed by Wald~\cite{Wald74}. The research demonstrates that the repulsion force is just great enough to prevent particles causing the destruction of event horizon to be captured by extremal Kerr-Newman black hole. Further research of Hubeny suggests that a near-extremal charged black hole can be overcharged~\cite{Hube99}. The work of Jacobson and Sotiriou for Kerr black hole further supports that near-extremal black hole can violate the weak cosmic censorship conjecture~\cite{JaSo09}. In this section, we use this method to check the validity of the weak cosmic censorship conjecture for $4D$ charged EGB black hole and consider the effect of the Guass-Bonnet coupling constant on the validity of weak cosmic censorship conjecture.

We shoot a test particle with rest mass $m$ and charge $\delta Q$ into the black hole on the radial direction. Due to the presence of electric repulsion force, the trajectory for the test charged particle in the $4D$ charged EGB spacetime is not geodesic. The equation of motion for the charged particle can be derived from the Lagrangian
\begin{equation}\label{Lagr}
  L= \frac{1}{2}mg_{\mu\nu}\frac{dx^\mu}{d\tau}\frac{dx^\nu}{d\tau} +\delta Q A_\mu\frac{dx^\mu}{d\tau}.
\end{equation}
From the Eular-Lagrangian equation, we can get the equation of motion, which is
\begin{equation}
  \frac{d^2x^\mu}{d\tau^2}+ \Gamma^\mu_{\alpha\beta}\frac{dx^\alpha}{d\tau}\frac{dx^\beta}{d\tau}=\frac{\delta Q}{m}F^\mu_{~~ \nu}\frac{dx^\nu}{d\tau}.
\end{equation}
Where $F^{\mu\nu}$ is the electromagnetic field tensor of the $4D$ charged EGB spacetime,
\begin{equation}\label{emtensor}
  F=\text{d}A=\frac{Q}{r^2}\text{dr}\wedge \text{dt}.
\end{equation}

Since we shoot the charged test particle into the black hole along the radial direction, the angular momentum of the particle is zero. The energy $\delta E$ of the particle is
\begin{equation}
  \delta E =-P_t=-\frac{\partial L}{\partial \dot{t}}= -mg_{00}^{~}\frac{dt}{d\tau}-\delta QA_t, \label{deltaEP}
\end{equation}
and the angular momentum of the particle
\begin{eqnarray}
% \nonumber to remove numbering (before each equation)
  P_\theta &=& \frac{\partial L}{\partial \dot{\theta}}=mg_{22}^{~}\frac{d\theta}{d\tau}=0, \\
   P_\phi&=& \frac{\partial L}{\partial \dot{\phi}}=mg_{33}^{~}\frac{d\phi}{d\tau}=0.
\end{eqnarray}

To check the validity of the weak cosmic censorship conjecture for the $4D$ charged EGB black hole, we first find the condition for particles to enter into the black hole, and then check whether the particle can destroy the event horizon of the black hole.

The four velocity of a massive particle is time-like and an unit vector, then, we have
\begin{equation}\label{fourV}
\begin{split}
  U^\mu U_\mu & =g_{\alpha\beta}\frac{dx^\alpha}{d\tau} \frac{dx^\beta}{d\tau}v \\
  &=\frac{1}{m^2}g^{\alpha\beta}(P_\alpha -\delta QA_{\alpha})( P_\beta-\delta QA_\beta) =-1.
  \end{split}
\end{equation}
Substituting the energy $\delta E$ and angular momentum $P_\theta$ and $P_\phi$ into the above equation, we have
\begin{equation}\label{EJM}
\begin{split}
  g^{00}\delta E^2+2g^{00}A_{t}\delta Q\delta E+g^{00}A_t^2\delta Q^2+g^{11}P_r^2 + m^2=0.
  \end{split}
\end{equation}
The above equation is a quadratic equation for $\delta E$. Solving the equation, we can get the energy of the charged particle,
\begin{equation}\label{EnergyP}
  \delta E=-A_t\delta Q- \frac{1}{g^{00}}\left[- g^{00}(g^{11}P_r^2+m^2)\right]^{\frac{1}{2}}.
\end{equation}

Where we have chosen the future directed solution $dt/d\tau>0$. Since the trajectory of a massive particle outside the event horizon of the $4D$ charged EGB black hole should be time-like and future directed. The future directed condition $dt/d\tau>0$ is equivalent to the following condition
\begin{equation}\label{Condition}
  \delta E>-A_t\delta Q.
\end{equation}
A particle falling into a black hole must cross the event horizon. Hence, the future directed condition for the charged particle at the event horizon implies
\begin{equation}\label{horicond}
  \delta E>\frac{Q}{r_\text{h}}\delta Q=\phi_\text{h} \delta Q.
\end{equation}
Thus, the condition for the particle to enter into the black hole is
\begin{equation}
  \delta E>\delta E_{\text{min}}\equiv\phi_{\text{h}}\delta Q. \label{Emin}
\end{equation}
This constraint guarantees that the test charged particle shooting into the black hole can fall into the event horizon. The constraint can also be derived from the null energy condition as in Ref.~\cite{JaSo09}.

On the other hand, to overcharge the black hole, the energy of the particle should not be too large. Thus, there should be an upper bound on the energy of the particle.
To first order, the condition to overcharge the black hole is
\begin{equation}
\begin{split}
  \Delta'_{\text{\text{min}}}=& -(M+\delta E)^2+(Q+\delta Q)^2+\alpha \\ =&-M^2+Q^2+\alpha -2M\delta E+2Q\delta Q>0,
  \end{split}
\end{equation}
which can be written as
\begin{equation}
  \delta E<\delta E_{\text{max}}=\frac{Q}{M}\delta Q-\frac{M^2-Q^2-\alpha}{2M}. \label{E1max}
\end{equation}
As long as the energy and charge of the test particle satisfy the two conditions Eq.~(\ref{Emin}) and Eq.~(\ref{E1max}), the horizon of the $4D$ charged EGB black hole can be destroyed and the weak cosmic censorship conjecture can be violated.

If the $4D$ charged EGB black hole starts out extremal, then we have $M^2-Q^2-\alpha=0$ and $\phi_{\text{h}}=Q/M $. Therefore, we have
 \begin{eqnarray}
 % \nonumber to remove numbering (before each equation)
   \delta E_{\text{max}} = \frac{Q}{M}\delta Q =\delta E_{\text{min}}.
 \end{eqnarray}
So $\delta E_{\text{min}}$ never less than $\delta E_{\text{max}}$. This indicates that particles causing the destruction of the event horizon just not be captured by the extremal $4D$ charged EGB black hole. This was claimed long ago by Wald for Kerr-Newman black hole~\cite{Wald74}.

However, if the black hole starts out very close to extremal, the event horizon of the black hole satisfys the inequality  $r_{\text{h}}=M+\sqrt{M^2-Q^2-\alpha}>M$. This inequality implies $\phi_{\text{h}}=Q/r_{\text{h}}<Q/M$. To first order, it is clear that there exist values of $\delta E$ satisfying both inequalities Eq.~(\ref{Emin}) and Eq.~(\ref{E1max}). So a near-extremal $4D$ charged EGB black hole can be overcharged and the weak cosmic censorship conjecture can be violated. Note that to first order, the Gauss-Bonnet coupling constant has no significant effect on the validity of the weak cosmic censorship conjecture.

Next, we consider the second order. To second order, the condition for overcharging the $4D$ charged EGB black hole is
\begin{equation}
  \Delta'_{\text{min}}= -(M+\delta E)^2+(Q+\delta Q)^2+\alpha>0. \label{E2minn}
\end{equation}
The condition to overcharge the $4D$ charged EGB black hole becomes that of Reissner-Nordstr\"om black hole for vanishing Gauss-Bonnet coupling constant. As previous research of Hubeny~\cite{Hube99} shows that extremal Reissner-Nordstr\"om black hole cannot be overcharged while near-extremal one can violate the weak cosmic censorship conjecture.

\begin{figure}
  \begin{center}
\subfigure[]{\includegraphics[width=3.2in]{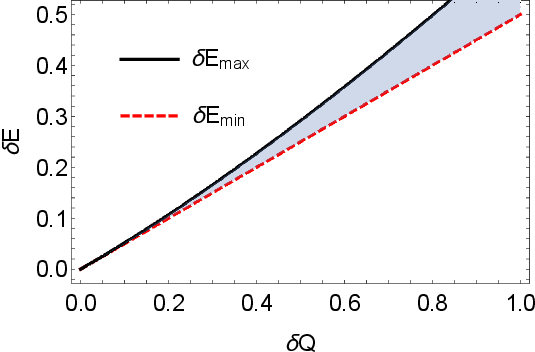}}
\subfigure[]{\includegraphics[width=3.2in]{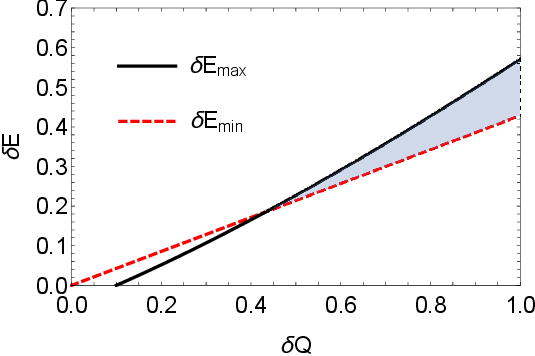}}
  \caption{(color online) Energy bounds for the charged particle $\delta E_{\text{max}}$ (black solid lines) and $\delta E_{\text{min}}$ ( red dashed lines) vs. charge of the particle $\delta Q$ for extremal and near-extremal $4D$ charged EGB black hole for positive coupling constant. Where we have choose the mass $M=2$ for the black hole, the Gauss-Bonnet coupling constant $\alpha =3$, and (a) charge of the extremal black hole $Q=1$, (b) charge of the near-extremal black hole $ Q=0.9$. The grey region is for $\delta E_{\text{max}}>\delta E_{\text{min}}$.  }\label{Figure1}
  \end{center}
\end{figure}

For positive coupling constant $\alpha$, the condition to destroy the event horizon of the black hole is
\begin{equation}
  \frac{(\delta E+M)^2}{\alpha}-\frac{(\delta Q+Q)^2}{\alpha}<1. \label{E2max}
\end{equation}
The above inequality describes a region below the upper branch of a hyperbola. It is clear from fig.~(\ref{Figure1}) that there exist particles with values of $\delta E$ and $\delta Q$ such that the two conditions (\ref{Emin}) and (\ref{E2max}) can be satisfied simultaneously both for extremal and near-extremal $4D$ charged EGB black holes. Hence, to second order, both extremal and near-extremal $4D$ charged EGB black holes can be destroyed for positive coupling constant.
We emphasize here, that due to the existence of the positive coupling constant, extremal charged EGB black hole can be overcharged, which is contrary to previous research that other extremal black holes cannot be overcharged or overspun by a test particle \cite{Wald74,JaSo09}.

For negative coupling constant $\alpha$, the condition to overcharge the black hole is
\begin{equation}
  \frac{(\delta Q+Q)^2}{|\alpha|}-\frac{(\delta E+M)^2}{|\alpha|}>1. \label{E3max}
\end{equation}
This gives the upper bound of energy for the particle
\begin{equation}
  \delta E<\delta E_{\text{max}}=\sqrt{(\delta Q+Q)^2-|\alpha|}-M. \label{E4max}
\end{equation}
Some algebra reveals that $\delta E_{\text{min}}$ never less than $\delta E_{\text{max}}$ for extremal $4D$ charged EGB black hole. So the event horizon of extremal $4D$ charged EGB black hole cannot be destroyed.

For near-extremal $4D$ charged EGB black hole, the validity of the weak cosmic censorship conjecture depends on the value of the Gauss-Bonnet coupling constant $\alpha$. If $0<-\alpha<<M$, the charge of the near-extremal black hole $Q\sim M$. The horizon of the black hole can be destroyed as that of Reissner-Nordstr\"om black hole. While for $-\alpha >>M$, the charge of the near-extremal black hole $Q>>M$. In this case, there is not any value of $\delta E$ satisfying both inequality (\ref{Emin}) and (\ref{E4max}), which shows that near-extremal black hole cannot be overcharged. Physically, this is because the electric repulsion force is too large for the charged particle to enter into the black hole.

Hence, to first order, extremal $4D$ charged EGB black hole cannot be overcharged; while the event horizon of near-extremal black hole can be destroyed. To second order, however, both extremal and near-extremal $4D$ charged EGB black hole can be destroyed for positive Gauss-Bonnet coupling constant; for negative Gauss-Bonnet coupling constant, extremal black hole cannot be overcharged and the validity of the weak cosmic censorship conjecture for near-extremal black hole depends on the value of the Gauss-Bonnet coupling constant.

\section{Comparing with other works}\label{6}

After our article appeared in arXiv a day later, Ying also posted a paper about weak cosmic censorship conjecture for the novel $4D$ charged EGB black hole on arXiv \cite{Ying20}. There are some similarities and differences between our work and Ying's research.

In our work, both charged scalar field scattering gedanken experiment and charged particle absorption experiment in the normal phase space were investigated. We considered the minimal value of the event horizon function $\Delta'_{\text{min}}$ to first order for scalar field scattering; and both first and second order for the minimal value of the event horizon function $\Delta'_{\text{min}}$ for charged particle absorption.

Ying considered particle absorption gedanken experiment both in the normal phase space and in the extended phase space, but the calculation of the minimal value of the event horizon function $f(r_{\text{min}})$ (in Ying's symbol) only to first order after the absorption of a charged particle.

In our work, we set the cosmological constant $\Lambda=0$, which makes it easier to calculate the minimal of the event horizon function $\Delta'_{\text{min}}$ to second order for particle injection. This leads to many interesting results, which cannot be derived from first order approximation, such as an extremal black hole can be overcharged for positive Gauss-Bonnet coupling constant. This is the main reason we only considered weak cosmic censorship conjecture in the normal phase space.

To first order of the minimal of the event horizon function, Ying's work should be the same with our research for particle absorption in the normal phase space, but Ying made a mistake in the estimation of the minimal of the event horizon function. After correcting the mistake, to first order, Ying's work is the same with our research.

In Ying's work, after the near-extremal black hole absorbing a charged particle with energy $E=q\Phi+| P^r(r_+)|$ and charge $q$, the minimal of the event horizon function is
\begin{equation}
\begin{split}
  f(r_{\text{min}}+dr_{\text{min}})=& \delta -\frac{2| P^r(r_+)|}{r_{\text{min}}
  +\frac{2\alpha}{r_{\text{min}}}(1-\delta)}\\
  &+ \frac{2Qq\epsilon'}{r_+^2(1-\epsilon'^2)+ \alpha(1-\delta)},
  \end{split}\label{Ying}
\end{equation}
where
\begin{equation}
 \delta\equiv f(r_{\text{min}})\leq0
\end{equation}
and
\begin{equation}
  r_{\text{min}}=r_+(1-\epsilon').
\end{equation}
Ying then claimed that the third term of Eq.~(\ref{Ying}) can be neglected in the near-extremal black hole case since $0<\epsilon'\ll1$, then Ying got the following result
\begin{equation}
  f(r_{\text{min}}+dr_{\text{min}})= \delta -\frac{2| P^r(r_+)|}{r_{\text{min}}+ \frac{2\alpha}{r_{\text{min}}}(1-\delta)}<0,
  \label{Ying1}
\end{equation}
and claimed that the weak cosmic censorship conjecture is valid for the near-extremal black hole. However, for vanishing Gauss-Bonnet coupling constant $\alpha\rightarrow0 $, Ying's conclusion contradicts with Hubeny's seminal work that a near-extremal Reissner-Nordstr\"om black hole can be overcharged in test particle approximation \cite{Hube99}. The reason for the contradiction is that Ying neglected the third term in Eq.~(\ref{Ying}).

The minimal of the event horizon function $\delta\equiv f(r_{\text{min}})$ for near-extremal black hole is of order $\epsilon'$, and we can choose injected particles with large charge $q$ but small radial momentum $0<|P^r(r_+)|<\epsilon'\ll1$.  The particle can cross the event horizon, since the energy of the particle $E=q\Phi+| P^r(r_+)|$ is large, too. In this case, after the absorption of the charged particle, the minimal of the event horizon function~(\ref{Ying}) can be positive. This leads to the result that weak cosmic censorship conjecture can be violated for a near-extremal black hole, and the result can be recovered to that of Reissner-Nordstr\"om black hole for vanishing Gauss-Bonnet coupling constant.

\section{Discussion and Conclusions}\label{7}

In this paper, we have explored the scenario of destroying the event horizon of the novel $4D$ charged EGB black hole with test field and test particle respectively. For the test charged field thought experiment, the result suggests that both extremal and near-extremal black hole cannot violate the weak cosmic censorship conjecture, and the result is independent of the Gauss-Bonnet coupling constant. For the test charged particle gedanken experiment,  to first order, extremal $4D$ charged EGB black hole cannot be overcharged and near-extremal black hole can be destroyed; to second order, however, both extremal and near-extremal $4D$ charged EGB black hole can be overcharged for positive Gauss-Bonnet coupling constant; for negative Gauss-Bonnet coupling constant, extremal black hole cannot be overcharged and the validity of the weak cosmic censorship conjecture for near-extremal black hole depends on the Gauss-Bonnet coupling constant.

\acknowledgments

We thank Yu-Xiao Liu, Shao-Wen Wei, Zi-Chao Lin and Wen-Bin Feng for many useful discussions. This work was supported in part by the National Natural Science Foundation of China (Grants No. 11875151, No. 11522541, and No. 11675064), and the Fundamental Research Funds for the Central Universities (Grants No. lzujbky-2019-it21).

\end{document}